\documentclass[12pt]{iopart}

\usepackage{amsfonts}

\newcommand{\Fo}{{\mathbb F}}

\newcommand{\be}{\begin{eqnarray}}
\newcommand{\ee}{\end{eqnarray}}

\begin{document}

\title [Grand canonical ensemble]
{Grand canonical ensemble in generalized thermostatistics}

\author{Jan Naudts and Erik Van der Straeten}
\address{Departement Fysica, Universiteit Antwerpen,\\
\small  Universiteitsplein 1, 2610 Antwerpen, Belgium}
\eads{\mailto{Jan.Naudts@ua.ac.be}, \mailto{Erik.VanderStraeten@ua.ac.be}}

\date {}


\begin{abstract}
We show that one can formulate a grand-canonical ensemble with uncertain number
of degrees of freedom without making specific assumptions about the
canonical ensemble.
Several choices of grand-canonical entropy functional are considered.
The ideal gas in non-extensive thermostatistics is discussed as an application.
\end{abstract}

\pacs {}

\section{Introduction}

In generalized thermostatistics some of the assumptions, which lay at the
basis of Gibbs' theory of the canonical ensemble, are not imposed. As a consequence,
the equilibrium probability distribution $p_{\beta,n}(x)$, with $x$ a point in
$n$-particle phase space $\Gamma_n$, is not necessarily of the Boltzmann-Gibbs
form
\be
p_{\beta,n}(x)=\frac 1{Z_n}\exp\left(-\beta H_n(x)\right).
\label {bg}
\ee
Here, $\beta$ is inverse temperature, $H_n(x)$ is the energy functional,
and $Z_n$ is the appropriate normalization factor.
On the contrary, $p_{\beta,n}(x)$ could be rather arbitrary, satisfying
some mild constraints of thermodynamic origin.

There may be several reasons why an experimental observation
yields a result which is not in agreement with (\ref {bg}).
It is not the goal of the present paper to discuss these reasons
in a systematic manner. It suffices to remark here that such
situations do occur frequently. Recent examples are \cite
{URGS01,LR01,JESK04,vz04}. Hence it is worthwhile to study generalized
thermostatistics from a theoretical point of view.

Many results of statistical mechanics do not depend on the actual choice of
equilibrium probability distribution for the canonical ensemble.
Often, the reason for this independence is the result known as
equivalence of ensembles, which is expected to hold for systems with many
degrees of freedom. The choice of (\ref {bg}) is usually convenient,
but not obligatory. In particular, deviations of (\ref {bg}) are expected
when the number of degrees of freedom is limited.
In the present paper is shown that a grand-canonical ensemble with an uncertain
number of degrees of freedom can be formulated without making specific assumptions
about the equilibrium distribution of the corresponding canonical ensemble.

Alternatives for the Boltzmann-Gibbs distribution (\ref {bg}) are highly
non-unique. Probability distributions proposed in the context of non-extensive
thermostatistics \cite {TC88} are suitable to serve as examples because these
distributions share many of the properties of (\ref {bg}). In particular, for the
distributions based on deformed exponential functions \cite {TC94,NJ02}
thermodynamic stability has been proved \cite {NJ03,NJ04b}. All these
distributions can be used as a basis to formulate a grand-canonical ensemble
in which the number of particles varies.

To be more specific we discuss the model of the ideal gas of
classical point particles. This model has been studied for the first time
in the context of non-extensive thermostatistics in \cite {PPT94}. However, this
and subsequent works \cite {YT99,AS99,MPP00,AMPP01} have treated the problem in a
canonical ensemble. In such an approach the phase spaces for $n=0,1,2,\cdots$
particles are all taken together into one infinite dimensional phase space.
This composite system is then treated in a canonical ensemble.
In a quantum-mechanical treatment the above procedure involves the construction of Fock
space. Then physical states occur which are superposition of states with
different number of particles. In the present paper the opposite point of view
is taken. The number of particles of the gas is fixed but unknown to the
experimenter. Hence, fluctuations in the number of particles should be
treated as a classical problem of lack of information involving an
ensemble of possibilities ($n=0,1,2,\cdots$), only one of which is realized.
On the other hand, the physical system contains a finite, possibly small,
number of particles. With appropriate boundary conditions, for example
those of \cite {vz04}, it can be correct to describe such a system as
being non-extensive.

Section 2 serves to fix notations. In Section 3 the probability $z_n$ that (part
of) the system contains $n$ particles is introduced. An
expression for the grand canonical entropy $S$ is proposed. In Section 4 these
probabilities $z_n$ are determined by means of a variational principle.
In Section 5 is shown that all usual thermodynamic relations are satisfied.
Sections 6, 7, and 8, treat non-extensive thermostatistics, and the ideal gas in
particular, as a special case. In Section 9 various alternatives are discussed.
The final section gives a summary of results and indicates possibilities for
further work. Appendix A deduces the generic form of the grand-canonical entropy.
Appendix B reproduces known results about the ideal gas in non-extensive
thermostatistics. Appendix C contains an alternative treatment of the ideal gas.

\section {A constant number of particles}

The equilibrium distribution $p_{\beta,n}$ is used to calculate average
energy $U_n(\beta)$ by
\be
U_n(\beta)=\langle H_n\rangle_\beta=\int_{\Gamma_n}{\rm
d}x\,p_{\beta,n}(x)H_n(x).
\label {udef}
\ee
Thermodynamic stability requires that $U_n(\beta)$ is a decreasing function of
$\beta$. There should exist an entropy function $S_n(U)$ with the property that
\be
\frac {{\rm d}\,}{{\rm d}U}S_n(U)=\beta \quad\hbox{ whenever }U=U_n(\beta)
\label {temp}
\ee
is satisfied.
$\beta$ is inverse temperature --- units are used in which $k_B=1$.
Given $p_{\beta,n}$, this entropy function can be obtained, up to
an integration constant, by integrating (\ref {temp}).

Consider e.g.~an ideal gas in dimension $\nu$. Its energy functional is given by
\be
H_n(v,q)=\sum_{j=1}^n\frac 12m|v_j|^2.
\ee
An example of an equilibrium distribution $p_{\beta,n}$, other than
Boltzmann-Gibbs, is a distribution of the form
\be
\begin {array} {rll}
p_{\beta,n}(x)
&\sim H_n(x)^{1-\nu n/2}
&\hbox { when }
(1-\epsilon)\frac {\nu n}{2\beta}\le H_n(x)\le (1+\epsilon)\frac {\nu n}{2\beta}\\
&=0
&\hbox{ otherwise}.
\end {array}
\label {altpdf}
\ee
with $\epsilon$ small but positive.
Such a distribution may be an appropriate tool to describe a gas
enclosed in a container and kept at constant pressure by means of a piston.
Energy fluctuations are then due to microscopic motions of the piston.
A short calculation gives
\be
U_n(\beta)=\frac {\nu n}{2\beta},
\label {idealU}
\ee
which is the same result as in case of a Boltzmann-Gibbs distribution.
A suitable choice of entropy is then
\be
S_n(U)=\frac {\nu n}2\ln(\beta_0 U)-\frac {\nu n}2\ln \frac {\nu n}2
+n\ln\frac V{L^\nu},
\label {idealS}
\ee
with $\beta_0$ given by
\be
\beta_0&=&\frac {2\pi m e L^2}{h^2}.
\ee
Here, $V$ is the volume and $L$ and $h$ are constants inserted for dimensional
reasons (we chose units in which $k_B=1$). In classical mechanics $h$ is a free
scaling parameter. In quantum mechanics it is Planck's constant and is fixed
by the ratio between energy and frequency of a photon.
The integration constant has been chosen in such a way that the standard
expression for entropy of an ideal gas is obtained.
As is well known, classical expressions of entropy, contrary to quantum
mechanical ones, become negative and diverge when average energy $U$ of the ideal
gas tends to zero. This non-physical behavior restricts the usefulness of the
ideal gas model to high energies.

\section{Grand canonical averages}

In the grand canonical ensemble a given number of particles $n$
occurs with probability $z_n$, normalized to
\be
\sum_{n=0}^\infty \frac {z_n}{c_n}=1.
\label {norm}
\ee
The numbers $c_n$ are usually taken equal to $1/n!$. This choice is
justified using the argument that permutation of particles
is a symmetry of the system.
Grand canonical averages are calculated using
\be
\langle f\rangle_\beta=\sum_{n=0}^\infty\frac {z_n}{c_n}
\int_{\Gamma_n}{\rm d}x\,p_{\beta,n}(x)f_n(x).
\ee
In particular, one has
\be
U(\beta,z)=\sum_{n=0}^\infty\frac {z_n}{c_n} U_n(\beta).
\ee
Average number of particles is given by
\be
N(z) = \sum_{n=1}^\infty \frac {z_n}{c_n}n,
\ee

However, grand canonical entropy $S(\beta,z)$ is not just an average of
canonical entropies $S_n(U_n(\beta))$ because additional entropy originates from
the uncertainty about the number of particles $n$.
It is shown in Appendix A that in equilibrium the grand canonical entropy
$S(\beta,z)$ must be of the form
\be
S(\beta,z)=\sum_n\frac {z_n}{c_n}S_n(U_n(\beta)) + \Fo(z),
\label {gcent}
\ee
with a function $\Fo(z)$ which depends on the probabilities $z_0,z_1,\cdots$
but may not depend on $\beta$. This motivates the definition
\be
S(\beta,z)=\sum_n\frac {z_n}{c_n}S_n(U_n(\beta))
-\sum_n\frac {z_n}{c_n}\ln z_n.
\label {gcentr}
\ee
Of course, other choices for the function $\Fo(z)$
are possible but there is no reason to take something else than
the most obvious expression for the entropy of the probabilities
$z_n$ with {\sl a priori} weights $c_n$.

\section{Grand canonical equilibrium}

The equilibrium values of the probabilities $z_n$ are obtained by minimization
of the free energy $U(\beta,z)-TS(\beta,z)$ under the constraint that the average
number of particles $N(z)$ has some given value. Equivalently, one maximizes
the Lagrange function
\be
{\cal L}(z)=S(\beta,z)-\beta U(\beta,z)
+\beta\mu N(z)-(\gamma-1)\sum_{n=0}^\infty\frac {z_n}{c_n}
\label {varprin}
\ee
with Lagrange parameters $\mu$ and $\gamma$, keeping $\beta$ fixed.
This leads to the set of equations
\be
\ln z_n=S_n(U_n(\beta))-\beta U_n(\beta)+\beta\mu n-\gamma.
\label {eqval}
\ee
Normalization gives
\be
\gamma=\ln\left[\sum_n\frac 1{c_n}\exp\{S_n(U_n(\beta))-\beta
U_n(\beta)+\beta\mu n\}\right].
\label {gamma}
\ee
One concludes that
\be
z_n=\frac 1Z e^{\beta\mu n}e^{S_n(U_n(\beta))-\beta U_n(\beta)}
\label {znstar}
\ee
with
\be
Z=e^\gamma=\sum_n \frac 1{c_n}e^{\beta\mu n}e^{S_n(U_n(\beta))-\beta
U_n(\beta)}.
\label {Zet}
\ee
The resulting equilibrium values of $S(\beta, z)$, $ U(\beta, z)$,
and $N(z)$ are denoted $S$, resp.~$U$ and $N$.

\section{Thermodynamical relations}

The grand-canonical potential $\Psi(\beta,\mu)$ is defined by \cite {CHB85}
\be
\Psi(\beta,\mu)=-\frac 1\beta \ln Z.
\label {thpot}
\ee
From the definition (\ref {gcentr}) follows, using (\ref {gamma}, \ref {znstar},
\ref {Zet}),\be
U=\Psi(\beta,\mu)+\frac 1\beta S+\mu N.
\label {gammatherm}
\ee
A short calculation, using (\ref {temp}), gives
\be
\frac{\partial\,}{\partial\beta}\Psi(\beta,\mu)=\frac 1{\beta^2} S,
\label {gammabeta}
\ee
and
\be
\frac{\partial\,}{\partial\mu}\Psi(\beta,\mu)=-N.
\label {gammamu}
\ee
Pressure $P$ is defined by
\be
P=-\frac{\partial\,}{\partial V}\Psi(\beta,\mu).
\ee
Hence it suffices to evaluate the partition sum $Z$ to obtain values for $N$,
$S$, $P$ and $U$ as a function of $\beta$ and $\mu$.

Note also that the following identities hold
\be
\frac 1\beta\frac {\partial S}{\partial\beta}
&=&\frac {\partial U}{\partial\beta}-\mu\frac {\partial N}{\partial\beta}\cr
\frac 1\beta\frac {\partial S}{\partial\mu}
&=&\frac {\partial U}{\partial\mu}-\mu\frac {\partial N}{\partial\mu}
\ee
(see Appendix A). Finally, from
\be
\frac {\partial^2\,}{\partial \beta\partial\mu}\ln Z=\frac
{\partial^2\,}{\partial \mu\partial\beta}\ln Z\ee
follows
\be
\frac {\partial S}{\partial\mu}=-\beta^2\frac{\partial N}{\partial\beta}.
\label {maxwell}
\ee

In case of an ideal gas, using (\ref {idealU}, \ref {idealS}), one obtains with
$c_n=1/n!$
\be
\Psi(\beta,\mu)=
-\frac V{L^\nu}\frac{\beta_0^{\nu/2}}{\beta^{1+\nu/2}}e^{\beta\mu-\nu/2}.
\ee
This implies
\be
N&=&\ln Z=\frac V{L^\nu}\left(
\frac{\beta_0}{\beta}\right)^{\nu/2}e^{\beta\mu-\nu/2}\cr
S&=&\left(1+\frac{\nu}{2}-\beta\mu\right)N\cr
U&=&\frac {\nu}{2\beta} N\cr
P&=&\frac 1\beta \frac NV.
\label {standard}
\ee

These results are of course not new. Point is that they can be derived
without assuming (\ref {bg}). The last of equations (\ref {standard}) is the
ideal gas law. The equation for the entropy can be written as
\be
S=N\ln\left(\left(\frac {\beta_0}{\beta}\right)^{\nu/2}\frac V{L^\nu}
\frac eN\right).
\ee
In this form it is known as the Sackur-Tetrode equation. This equation
agrees well with experimental data --- see \cite {MQD00}, Section 5-3.
The heath capacity $C_V$ evaluates to
\be
C_V=-\frac {\beta}N\frac {\partial S}{\partial\beta}=\frac {\nu}2
+\left(\beta\mu-\frac {\nu}2\right)^2.
\ee
The first term is half the constant of Dulong and Petit.
The latter term is the contribution due to the temperature dependence of the
number of particles.

\section{Non-extensive thermostatistics}

In recent literature on non-extensive thermostatistics \cite {TMP98,MNPP00}
average energy $U_n(\beta)$ is calculated using the probability distribution
\be
p_{\beta,n}(x)=\frac 1{Z_n(\beta)}
\left[1-(1-q)\beta_q[H_n(x)-U_n(\beta)]\right]_+^{q/(1-q)}.
\label {pdfabe}
\ee
Note that this is the so-called escort probability distribution \cite {TMP98}.
In particular, (\ref {udef}) holds for this distribution.
In (\ref {pdfabe}) $Z_n$ is the normalization
\be
Z_n(\beta)=\int_{\Gamma_n}{\rm d}x\,
\left[1-(1-q)\beta_q[H_n(x)-U_n(\beta)]\right]_+^{q/(1-q)}
\ee
and $\beta_q$ is related to inverse temperature $\beta$ by
\be
\beta_q=\beta \left(\frac mh\right)^{\nu n}\int_{\Gamma_n}{\rm d}x\,p_{\beta,n}(x)^{1/q}
=\beta\left[\left(\frac mh\right)^{\nu n}Z_n(\beta)\right]^{-(1-q)/q}.
\label {betaqdef}
\ee
The notation $[u]_+=\max\{0,u\}$ is used.
Note that the expression for $U_n(\beta)$, based on (\ref  {pdfabe}),
is self-referential. In general, there is no guarantee that this implicit
equation yields a solution $U_n(\beta)$. In addition, this solution should be a
decreasing function of $\beta$.
Entropy $S_n(U)$ is given by
\be
S_n(U_n(\beta))=\frac q{1-q}\left(\left[\left(\frac mh\right)^{\nu n} Z_n(\beta)\right]^{(1-q)/q}-1\right).
\label {neS}
\ee
With this definition (\ref {temp}) is satisfied. Hence the grand-canonical
ensemble can be introduced along the lines of Sections 3, 4, 5.

For the ideal gas, one can show that the implicit equation for $U_n(\beta)$ has a
solution (See \cite {AS99,MPP00};
For sake of completeness the calculation is reproduced in Appendix B).
It is given by (\ref {idealU}), with $\beta$ replaced by $\beta_q$.
A problem that now arises is that $\beta_q$ is an increasing function
of $\beta$ only if the condition
\be
\frac {1-q}q\frac {\nu n}2<1.
\label {qreq}
\ee
is satisfied (see Appendix B).
We need this condition for stability reasons: energy should be an increasing function
of temperature. Hence (\ref {qreq}) must hold. But clearly, for fixed $q$ the condition
is always violated if $n$ becomes large ($0<q<1$ is assumed the whole time).
Hence our approach, involving sums over al $n$, is not valid for the ideal gas
in combination with the non-extensive probability distribution (\ref {pdfabe}).

In \cite {AMPP01} the assumption is made that $\beta_q=\beta$, instead of
(\ref {betaqdef}). Then the appropriate entropy for which (\ref {temp}) holds
is
\be
S_n(U_n(\beta))=\ln\left[\left(\frac mh\right)^{\nu n} Z_n(\beta)\right].
\label {neS2}
\ee
As noted in \cite {AMPP01}, equipartition now holds in
its original form (\ref {idealU}). Expression (\ref {neS2}) reduces to (\ref {idealS}).
Hence the grand-canonical description of the ideal gas with fluctuating number of
particles is described by (\ref {standard}). This shows that not only
(\ref {altpdf}) but also (\ref {pdfabe}), with $\beta_q=\beta$,
cannot be distinguished from (\ref {bg}).
The question whether $\beta$ should equal $\beta_q$ or should be given by (\ref {betaqdef})
is intricate and is related to the choice of canonical and microcanonical entropy.
See the discussion in \cite {AMPP02} and the comment on this paper made in \cite {TR03}.
Further investigations are needed to clarify this point.

\section {Alternative treatment}

In Section 3  the assumption is made that the grand-canonical entropy has two
contributions, the average of entropies at fixed number of particles $n$ and a
contribution due to fluctuations in $n$. For the latter the expression of Shannon
information has been used. This is the most logical choice in our approach
because the variational principle is used only to deduce the ensemble with
variable number of particles from the ensemble with fixed number of particles.
Quite often the equilibrium state at constant $n$ is already obtained
by means of a variational principle. In that case our approach consists
of two subsequent optimizations of an entropy functional.
One can prefer to do this optimization in one single step. This yields of course
the same results. However, one could then argue that Shannon information is
inappropriate to describe fluctuations in the number of particles in case
another type of entropy functional is already used at the level of a fixed number
of particles.

An example of an entropy functional, which can replace (\ref {gcentr}), is
\be
S(p)
&=&\sum_{n=0}^\infty \frac 1{c_n}\frac 1{(2-q_n)(q_n-1)}
\int_{\Gamma_n}{\rm d}x\,p_n(x)
\left[\left(\frac {h^{\nu n}L^{\nu n}}{m^{\nu n}V^n}p_n(x)\right)^{1-q_n}-1\right].\cr
& &
\label {altent}
\ee
Each term in this sum is essentially the entropy functional advocated by Tsallis
\cite {TC88}, with $q_n$ replaced by $2-q_n$ --- see \cite {NJ04b} for an
explanation why this change is made. Note that, unlike in section 6, the
parameter $q$ is made $n$-dependent and is assumed to be larger than 1. This
assumption is made just to show that it is also possible.
The entropy functional (\ref {altent}) is maximized under the constraints that
average energy $U$ and average particle number $N$ should have specified values.
This optimization involves Lagrange multipliers $\beta$, $\mu$ and $\gamma$, much
like in (\ref {varprin}). The resulting equilibrium probability distribution is
\be
p_n(x)=\left(\frac {mV^{1/\nu}}{hL}\right)^{\nu n}
\exp_{q_n}\bigg(-\frac 1{2-q_n}-\gamma-\beta H_n(x)+\beta\mu n\bigg)\cr
& &
\label {alteq}
\ee
with the deformed exponential function defined by \cite {TC94,NJ02}
\be
\exp_q(u)=[1+(1-q)u]_+^{1/(1-q)}.
\ee
The requirement of normalization reads
\be
1=\sum_{n=0}^\infty\frac 1{c_n}\int_{\Gamma_n}{\rm d}x\,p_n(x).
\ee
The result (\ref {alteq}) is {\sl not} of the form
\be
p_n(x)=z_np_{\beta,n}(x),
\ee
where $p_{\beta,n}(x)$ are equilibrium distributions at fixed number of
particles. However, one can show that $p_n(x)$, as given by (\ref {alteq}),
satisfies all thermodynamic relations discussed in Section 5. See Appendix C.
Hence, from a thermodynamic point of view
this alternative treatment is fully acceptable. Technically, this choice of
entropy functional is more difficult than (\ref {gcentr}) because an analytic
expression for normalization $\gamma$, similar to (\ref {gamma}), is missing.

\section{Discussion}

In the first part of this paper we have given a description of a gas with
a fluctuating number of particles $n$ without specifying the equilibrium
distribution at fixed $n$. Motivation for making this rather trivial
extension of the standard grand-canonical description is that in many
experiments the conditions for a constant temperature ensemble at fixed $n$ are
not satisfied. In the extended theory standard thermodynamic relations
are still satisfied provided that entropy and energy at fixed $n$ satisfy the
well-known temperature relation (\ref {temp}). A simple example is the
ideal gas model for which standard results like the ideal gas law or the
Sackur-Tetrode equation are obtained without assuming a Boltzmann-Gibbs
distribution.

In the last part of the paper non-extensive thermostatistics
is considered as a special case of generalized thermodynamics.
The distributions of non-extensive thermostatistics are
characterized by the fact that the probability of large energies
does not decrease exponentially as is the case with Boltzmann-Gibbs.
Rather they go either with a powerlaw (called the $q>1$-case) or
they have a cutoff energy above which configurations have vanishing
probability ($q<1$-case). Point of discussion in non-extensive thermostatistics
is how to define inverse temperature $\beta$. Inherently related is the
definition of thermodynamic entropy, which is obtained by integrating
the thermodynamic relation (\ref {temp}). Following \cite {TMP98,MNPP00},
the parameter $\beta_q$ occuring in the equilibrium probability distribution (\ref {pdfabe})
is {\sl not} $\beta$ but is related to it by (\ref {betaqdef}).
For the ideal gas, (\ref {betaqdef}) implies that the non-extensivity parameter $q$
must go to 1 as the number of particles $n$ becomes large. But, our treatment
of the grand-canonical ensemble involves sums over all values of $n$. It is
therefore not applicable in this case, except when $q=1$.

On the other hand, \cite {AMPP01} made the assumption that $\beta_q=\beta$.
Then the equipartition law ($(1/2)k_BT$ of energy per degree of freedom) still holds.
With an appropriate choice of entropy $S_n(U_n)$ the thermodynamic description at constant number
of particles is identical with that of the extensive case. As a consequence, also the
grand-canonical descriptions coincide.

Our treatment of the grand-canonical ensemble is kept as simple as possible.
The assumption is made that the fluctuation of the number of particles
is well described by a maximum-entropy principle involving Shannon's measure
of information. However, many alternatives can be considered. One of them
is discussed in section 7.

The present work is only a first step in describing real gases using
generalized thermostatistics. The logically next step is a study of the
virial expansion. One can expect that the virial coefficients will
depend on the detailed form of the canonical equilibrium distribution.

\appendix
\setcounter{section}{0}

\section{Form of the grand canonical entropy}

Here we show that in equilibrium the grand canonical entropy
must be of the form (\ref {gcent}).
From (\ref {varprin}) it follows that
\be
\gamma&=&1+c_n\frac{\partial\,}{\partial z_n}\left[
S-\beta U+\beta\mu N\right].
\ee
Now, $S$ and $U$
depend on $\mu$ only via $z_n$. As a consequence, one has
\be
\frac{\partial S}{\partial\mu}-\beta \frac{\partial U}{\partial\mu}
&=&\sum_n\frac{\partial\,}{\partial z_n}\left[S-\beta U\right]
\frac{\partial z_n}{\partial\mu}\cr
&=&\sum_n\frac 1{c_n}(\gamma-1-\beta\mu n)\frac{\partial z_n}{\partial\mu}\cr
&=&-\beta\mu \frac{\partial N}{\partial\mu}.
\label {tdrel1}
\ee
To obtain the latter we used that from normalization follows that
\be
0=\sum_n\frac 1{c_n}\frac{\partial z_n}{\partial\mu}.
\ee

From (\ref {tdrel1}) follows that the potential $\Psi(\beta,\mu)$ of the grand
canonical ensemble satisfies the thermodynamic relation
\be
\frac {\partial\,}{\partial\mu}\Psi(\beta,\mu)=-N.
\ee
This shows that the parameter $\mu$ is indeed the chemical potential.

Next consider temperature dependence.
From (\ref {thpot}) follows
\be
\frac {\partial\,}{\partial\beta}\Psi(\beta,\mu)
&=&\frac 1{\beta^2}S-\frac 1\beta\frac {\partial S}{\partial\beta}
+\frac {\partial U}{\partial\beta}-\mu\frac {\partial N}{\partial\beta}.
\ee
The thermodynamic relation that should hold is
\be
\frac {\partial\,}{\partial\beta}\Psi(\beta,\mu)
&=&\frac 1{\beta^2}S.
\ee
Hence grand canonical entropy should be such that
\be
\frac 1\beta\frac {\partial S}{\partial\beta}
&=&\frac {\partial U}{\partial\beta}-\mu\frac {\partial N}{\partial\beta}\cr
&=&\sum_n\frac {z_n}{c_n}\frac {{\rm d} \,}{{\rm d}\beta}U_n(\beta)
+\sum_n\frac {\partial z_n}{\partial\beta}\frac{\partial\,}{\partial z_n}
(U-\mu N)\cr
&=&\sum_n\frac {z_n}{c_n}\frac {{\rm d} \,}{{\rm d}\beta}U_n(\beta)
+\frac 1\beta\sum_n\frac {\partial z_n}{\partial\beta}
\left(\frac{\partial S}{\partial z_n}-\frac{\gamma-1}{c_n}\right)\cr
&=&\sum_n\frac {z_n}{c_n}\frac {{\rm d} \,}{{\rm d}\beta}U_n(\beta)
+\frac 1\beta\sum_n
\frac{\partial S}{\partial z_n}
\frac {\partial z_n}{\partial\beta}.
\ee
Using (\ref {temp}) the requirement becomes
\be
\frac {\partial S}{\partial\beta}
&=&\sum_n\frac {z_n}{c_n}\frac {{\rm d} \,}{{\rm d}\beta}S_n(\beta)
+\sum_n
\frac{\partial S}{\partial z_n}
\frac {\partial z_n}{\partial\beta}.
\label {req}
\ee
This means that entropy should be the sum of two contributions,
some function of the $z_n$ plus average canonical entropy.

\section{Non-extensive distributions}

Here we check that equipartition $\beta_q U_n(\beta)=\nu n/2$ is a solution of
the implicit equation
\be
U_n(\beta)&=&\int_{\Gamma_n}{\rm d}x\,p_{\beta,n}(x)H_n(x)\cr
&=&\frac 1{Z_n(\beta)}\int_{\Gamma_n}{\rm d}x\,
\left[1-(1-q)\beta_q[H_n(x)-U_n(\beta)]\right]_+^{q/(1-q)}H_n(x)\cr
& &
\ee
in case $H_n(x)$ is the Hamiltonian of an ideal gas
(The parameter $q$ is assumed to lie between 0 and 1).
Equivalently, we show that
\be
Z_n(\beta)&=&\int_{\Gamma_n}{\rm d}x\,
\left[1-(1-q)\beta_q[H_n(x)-U_n(\beta)]\right]_+^{1/(1-q)}.
\label {Zrel}
\ee

Let $\zeta(n)$ denote the volume of the unit sphere in $n$ dimensions.
It satisfies
\be
\zeta(n)=\frac {\pi^{n/2}}{\Gamma(n/2+1)},
\ee
This is used to calculate
\be
& &\int_{\Gamma_n}{\rm d}x\,
\left[1-(1-q)\beta_q[H_n(x)-U_n(\beta)]\right]_+^\alpha\cr
&=&V^n\zeta(\nu n)\int_0^{+\infty}r^{\nu n-1}{\rm d}r\,
\left[1-(1-q)\beta_q\frac m2r^2+(1-q)\frac {\nu n}2\right]_+^\alpha\cr
&=&V^n\zeta(\nu n)
\left(1+(1-q)\frac {\nu n}2\right)^{\alpha+\nu n/2}\cr
& &\times
\left[(1-q)\beta_q\frac m2\right]^{-\nu n/2}
\frac {\Gamma(\nu n/2)\Gamma(\alpha+1)}{2\Gamma(\alpha+1+\nu n/2)}.
\label {B4}
\ee
It is straightforward to verify that this quantity
has the same value for $\alpha=q/(1-q)$ and for $\alpha=1/(1-q)$.
Hence (\ref {Zrel}) holds.

From (\ref {B4}) follows that $Z_n(\beta)$ is proportional to $\beta_q^{-\nu n/2}$.
In combination with (\ref {betaqdef}) this implies that
$\beta$ is proportional to $\beta_q^{1-(1-q)\nu n/2q}$.
But $\beta_q$ should be an increasing function of $\beta$. This leads to
requirement (\ref {qreq}).

\section{Alternative choice of grand-canon\-ical entropy functional}

Here, we show that the choice of entropy functional (\ref {altent})
is acceptable from a thermodynamic point of view.
The grand-canonical potential $\Psi$ of generalized thermostatistics can be
defined by
\be
-\beta\Psi=S-\beta U+\beta\mu N.
\label {gcpot}
\ee
Inverse temperature $\beta$ in the grand-canonical ensemble is defined by
\be
S=\beta^2\frac {\partial\Psi}{\partial\beta}.
\label {gctemp}
\ee
Note that (\ref {gcpot}) implies that
\be
\beta^2\frac {\partial\Psi}{\partial\beta}
=S-\beta\frac {\partial S}{\partial\beta}+\beta^2\frac{\partial
U}{\partial\beta}-\beta^2\mu\frac{\partial N}{\partial\beta}.
\ee
Hence, to show that (\ref {gctemp}) holds, it suffices to show that
\be
\frac {\partial S}{\partial\beta}=\beta\frac{\partial
U}{\partial\beta}-\beta\mu\frac{\partial N}{\partial\beta}
\label {tr1}
\ee
One calculates
\be
\frac {\partial S}{\partial\beta}
&=&\sum_{n=0}^\infty \frac 1{c_n}\frac 1{(2-q_n)(q_n-1)}\cr
& &\times
\int_{\Gamma_n}{\rm d}x\,\left[
(2-q_n)\left(\frac {h^{\nu n}L^{\nu n}}{m^{\nu
n}V^n}p_n(x)\right)^{1-q_n}-1\right]\frac{\partial\,}{\partial\beta}p_n(x)\cr
&=&\sum_{n=0}^\infty \frac 1{c_n}
\int_{\Gamma_n}{\rm d}x\,
(\gamma+\beta H_n(x)-\beta \mu n)
\frac{\partial\,}{\partial\beta}p_n(x)\cr
&=&\beta\frac{\partial U}{\partial\beta}+\beta\mu\frac{\partial
N}{\partial\beta}.
\ee
This shows (\ref {tr1}) and implies (\ref {gctemp}).

Chemical potential $\mu$ is correctly defined if
\be
N=-\frac {\partial\Psi}{\partial\mu}.
\ee
It suffices to show that
\be
\frac 1\beta\frac {\partial S}{\partial\mu}
&=&\frac {\partial U}{\partial\mu}-\mu\frac {\partial N}{\partial\mu}.
\label {gcchem}
\ee
One calculates
\be
\frac 1\beta\frac {\partial S}{\partial\mu}
&=&\sum_{n=0}^\infty \frac 1{c_n}\frac 1{(2-q_n)(q_n-1)}\cr
& &\times
\int_{\Gamma_n}{\rm d}x\,
\left[(2-q_n)\left(\frac {h^{\nu n}L^{\nu n}}{m^{\nu n}V^n}p_n(x)\right)^{1-q_n}
-1\right]\frac 1\beta \frac{\partial\,}{\partial\mu}p_n(x)\cr
&=&\sum_{n=0}^\infty \frac 1{c_n}\int_{\Gamma_n}{\rm d}x\,
\left[\gamma+\beta H_n(x)-\beta\mu n\right]
\frac 1\beta \frac{\partial\,}{\partial\mu}p_n(x)\cr
&=&\frac {\partial U}{\partial\mu}-\mu\frac {\partial N}{\partial\mu}.
\ee
This shows (\ref {gcchem}).
Finally, (\ref {maxwell}) follows from
\be
\frac {\partial^2\,}{\partial\mu\partial\beta}\Psi=\frac
{\partial^2\,}{\partial\beta\partial\mu}\Psi.
\ee



\section*{References}
\begin{thebibliography}{99}

\raggedright

\bibitem {URGS01} A. Upadhyaya, J.-P. Rieu, J.A. Glazier, Y. Sawada,
{\sl Anomalous diffusion and non-Gaussian velocity distribution of Hydra cells
in cellular aggregates,} Physica A{\bf 293}, 549-558 (2001).

\bibitem {LR01} V. Latora, A. Rapisarda,
{\sl Dynamical quasi-stationary states in a system with long-range forces,}
Chaos Solitons and Fractals 13, 401-406 (2001).

\bibitem{JESK04} J. Jersblad, H. Ellmann, K. St\o chkel, A. Kastberg,
{\sl Non-gaussian velocity distributions in optical lattices,}
arXiv:cond-mat/0304358, Phys. Rev. A{\bf 69}, 013410 (2004).

\bibitem {vz04} J.S. van Zon, J. Kreft, D.I. Goldman, D. Miracle, J.B. Swift,
H.L. Swinney, {\sl Crucial role of sidewalls in velocity distributions in
quasi-2D granular gases,} arXiv:cond-mat/0405044, Phys. Rev. E{\bf 70}, 040301 (2004).

\bibitem {TC88}  C. Tsallis, {\sl Possible Generalization of
Boltzmann-Gibbs Statistics,}
J. Stat. Phys. {\bf 52}, 479-487 (1988).

\bibitem{TC94}C. Tsallis,
{\sl What are the numbers that experiments provide?}
Quimica Nova {\bf 17}, 468 (1994).

\bibitem {NJ02} J. Naudts, {\sl Deformed exponentials and logarithms
in generalized thermostatistics,} arXiv:cond-mat/0203489,
Physica A{\bf 316}, 1-12 (2002).

\bibitem {NJ03} J. Naudts, {\sl Generalized thermostatistics and mean field
theory,} arXiv:cond-mat/0211444v6, Physica A{\bf 332}, 279-300 (2004).

\bibitem {NJ04b} J. Naudts, {\sl Generalized thermostatistics based on
deformed exponential and logarithmic functions,} arXiv:cond-mat/0311438,
Physica A{\bf 340}, 32-40 (2004).

\bibitem {PPT94} A.R. Plastino, A. Plastino, C. Tsallis,
{\sl The classical $N$-body problem within a generalized
statistical mechanics,} J. Phys. A{\bf 27}, 5707-5714 (1994).

\bibitem {YT99} T. Yamano,
{\sl $N$-body classical systems with $r^\alpha$ interparticle potential and
two-particle correlation function using Tsallis statistics,}
Phys. Lett. A{\bf 264}, 276-282 (1999).

\bibitem {AS99} S. Abe, {\sl Thermodynamic limit of a classical gas in
nonextensivestatistical mechanics: negative specific heath and polytropism,}
Phys. Lett. A{\bf 263}, 424-429 (1999); A{\bf 267}, 456-457 (2000).

\bibitem {MPP00} S. Martinez, F. Pennini, A. Plastino,
{\sl Equipartition and viral theorems in a nonextensive optimal Lagrange
multipliers scenario,}
Phys. Lett. A{\bf 278}, 47-52 (2000).

\bibitem {AMPP01} S. Abe, S. Martinez, F. Pennini, A. Plastino,
{\sl Classical gas in nonextensive optimal Lagrange multipliers formalism,}
Phys. Lett. A{\bf 278}(5), 249-254 (2001).

\bibitem {CHB85} H.B. Callen, {\sl Thermodynamics and an introduction to
thermostatistics,} 2nd ed. (John Wiley \& Sons, Inc., 1985)

\bibitem {MQD00} D. McQuarrie, {\sl Statistical Mechanics}
(University Science Books, 2000)

\bibitem {TMP98} C. Tsallis, R.S. Mendes, A.R. Plastino,
{\sl The role of constraints within generalized nonextensive
statistics,} Physica A{\bf 261}, 543-554 (1998).

\bibitem {MNPP00} S. Martinez, F. Nicol\'as, F. Pennini, A. Plastino, {\sl
Tsallis' entropy maximization procedure revisited,} Physica A{\bf 286}, 489-502
(2000).

\bibitem {AMPP02} S. Abe, S. Martinez, F. Pennini, A. Plastino,
{\sl Nonextensive thermodynamic relations,}
Phys. Lett. A{\bf 281}, 126-130 (2001).

\bibitem {TR03} R. Toral, {\sl On the definition of physical temperature and pressure for
nonextensive thermostatistics,} Physica A{\bf 317}, 209-212 (2003).

\end{thebibliography}
\end{document}